\documentclass[RNAAS]{aastex62}

\begin{document}

\title{An Improved Orbital Period for GY Cancri Based on Two K2 Campaigns}

\correspondingauthor{Peter Garnavich}
\email{pgarnavi@nd.edu}

\author{Luis Alberto Ca\~{n}izares}
\affiliation{School of Physical Sciences, Dublin City University, Dublin 9, Ireland}
\affiliation{Physics Department, University of Notre Dame, Notre Dame, IN 46566}

\author{Peter Garnavich}
\affiliation{Physics Department, University of Notre Dame, Notre Dame, IN 46566}

\author{Colin Littlefield}
\affiliation{Physics Department, University of Notre Dame, Notre Dame, IN 46566}

\author{Joshua Pepper}
\affiliation{Physics Department, Lehigh University, Bethlehem, PA 18015}

\author{Allyson Bieryla}
\affiliation{Harvard-Smithsonian Center for Astrophysics, Cambridge, MA 02138}

\author{Supachai Awiphan}
\affiliation{National Astronomical Research Institute of Thailand, 260, Moo 4, T. Donkaew, A. Mae Rim, Chiang Mai, 50180, Thailand}

\author{Siramas Komonjinda}
\affiliation{Department of Physics and Materials Science, Faculty of Science, Chiang Mai University, 239, Huay Keaw Road, T. Suthep, A. Muang, Chiang Mai, 50200, Thailand}


\section{} 

GY~Cnc is a deeply eclipsing cataclysmic variable star that has shown several dwarf nova (DN) outbursts \citep{kato02, feline05} and is an X-ray source \citep{bade98}. The variable was continuously observed by the K2/Kepler satellite in Campaign 5 (C05) for 75 days during 2015. The star was again observed in 2017/2018 for 80 consecutive days during Campaign 16 (C16). Both sets of observations were made in the short cadence mode (1-minute exposures) and the standard aperture was used to extract the flux during C05. For C16 the star was far from the optical axis and a special aperture was created to collect a majority of the light in the K2 pixel file. The K2 light curves did not include a DN outburst, but the system varied on a time-scale of days by a factor of two outside of eclipse during C05 and nearly a factor of two over C16.

To determine times of minima, eclipses in the K2 photometry were fit using a quadratic function. 419 well-observed eclipses were measured over C5 and 446 timings were determined in C16. The root-mean-square (RMS) scatter in the K2 timing residuals was 21~s. We compared the times of minimum in K2 to the ephemeris of \citet{feline05} and found a significant offset. Therefore, a new ephemeris was calculated combining the K2 data, published timings and moderately high-cadence observations in the American Association of Variable Star Observers (AAVSO) database contributed by observers Tonny Vanmunster, Etienne Morelle, Francois Teyssier, and Richard Stanton. In addition, the KELT Follow-Up Network observed three eclipses during C16, with KeplerCam on the 1.2-m telescope at the Fred Lawrence Whipple Observatory, and the TRT-TNO telescope in Thailand, and these observations confirm the K2 timing measurements.

\begin{figure}[h!]
\begin{center}
\includegraphics[width=\textwidth]{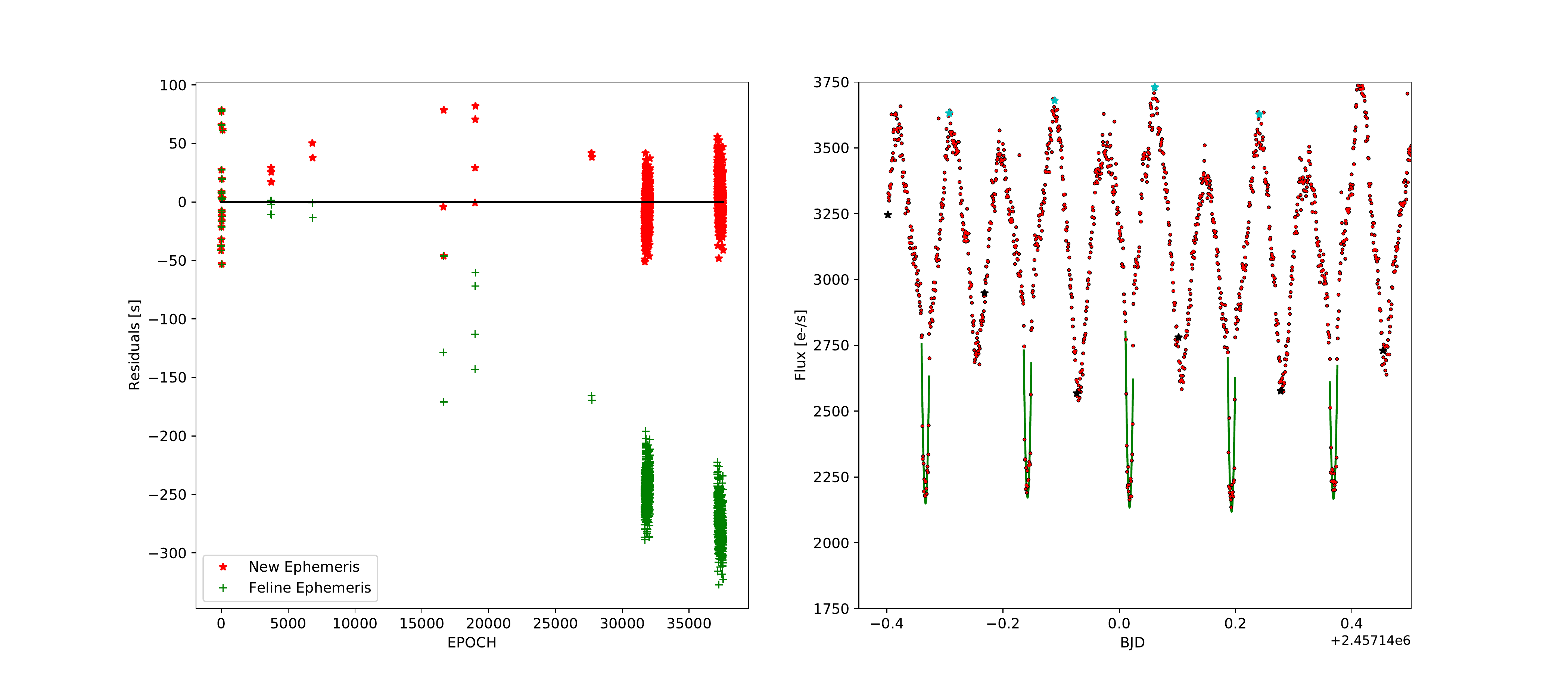}
\caption{ {\bf Left:} The eclipse timing residuals (red points) obtained from the ephemeris given by Eqn \ref{eqn:ephemeris}, derived from previously published observations and eclipses observed by K2 during C5 and C16. The corresponding \citet{feline05} ephemeris residuals are displayed as green points. {\bf Right:} A small portion of the K2 C05 light curve showing the eclipse timing estimates in green. The orbital modulation is strong, so we estimated long-term variations of the system by identifying the peak brightness in each orbit (shown as blue dots in the plot) \label{fig:1}}
\end{center}
\end{figure}

The combined eclipse timings gives a linear orbital ephemeris in barycentric Julian days of:

\begin{equation}\label{eqn:ephemeris}
 \begin{centering}
BJD = 2451586.212797(40) + 0.1754424121(12)\; E\;\; ,
  \end{centering}
\end{equation}
where $E$ is the epoch number counting from 2000 February 11, and the uncertainties are shown in the parentheses for the final digits of the period and initial epoch. 
The new proposed orbital period of GY~Cnc is $P_{orb} = 0.1754424121(12)$ days, or 4 hrs 12 mins 38.2244 s which adjusts by 7.5~ms the period presented by \cite{feline05}. While the revision is small, the accumulated errors amounted to 300~s by C16 (see Fig~\ref{fig:1}). The extended baseline of eclipse timings does not show the need for a quadratic term, suggesting that there is currently no detectable period derivative.

We find a correlation between the eclipse timing residuals and the brightness of the system. Over each orbit we identify the peak flux in the K2 data and use that as an indicator of the system luminosity. For both C05 and C16, we find the eclipse arrives as much as 15~s earlier than average when the system is at its faintest. When it was at its brightest, the eclipses occurred as much as 15~s late compared to the ephemeris. Eclipse phase variations of 3\%\ between quiescence and DN outbursts have been documented in Kepler observations of V447~Lyr \citep{ramsay12}. The quiescent phase shifts of $\sim 0.2$\%\ seen here in GY~Cnc may result from changes in the hotspot brightness due to small variations in the accretion rate.

In conclusion, the recent observations of GY~Cnc by K2, the AAVSO, and KELT have refined the orbital period of GY Cnc and improved its ephemeris which had accumulated an error of about 300~s over 18~years. The revised period is found to be 4 hrs 12 mins 38.2244 s. A quadratic term was not found to be significant indicating that there is currently no detectable orbital period derivative. We observed a correlation between the quiescent system brightness and the eclipse timing residuals that are as large as $\pm 15$~s in the K2 data.

\facilities{K2, AAVSO, KELT}


\end{document}